\documentclass[epj,final]{svjour}

\usepackage{amsfonts,amssymb}

\usepackage[dvips]{epsfig}

\date{\today}
\begin{document}   
\title{On the %
origin of surface states in a correlated local-moment film}
\author{R.~Schiller\inst{1} \and W.~M\"uller\inst{1} \and 
        W.~Nolting\inst{1}}
\institute{Humboldt-Universit\"at zu Berlin, Lehrstuhl
  Festk\"orpertheorie, Invalidenstra{\ss}e 110, D-10115 Berlin, Germany\\
  phone: +49 (30) 2093 7640 ~~fax: +49 (30) 2093 7643 
  \email{nolting@physik.hu-berlin.de}
  }

\abstract{    
The electronic quasiparticle structure of a ferromagnetic local moment film is
investigated within the framework of the s-f model. For the special case of a
single electron in an otherwise empty energy band being exchange coupled to a
fully ordered localised spin system the problem can be solved exactly and, 
for the
spin-$\downarrow$ electron, some marked correlation effects can be found. We
extend our model to incorporate the influence of the
surface on the electronic structure. 
Therefore we modify the hopping integrals in the vicinity of the
surface. This leads to the existence of surface states, both for the
spin-$\uparrow$ and the spin-$\downarrow$ spectral density of states. The
interplay between the modification of the hopping integrals and the existence
of surface states and correlation effects is discussed in detail.
\PACS{
  {75.70.-i}{Magnetic films and multilayers}\and
  {73.20.At}{Surface states, band structure, electron density of states}\and
  {75.50.Pp}{Magnetic semiconductors}
}
}

\maketitle

\titlerunning{Surface states in a local-moment film}

\section{Introduction}
In the recent past the s-f (or s-d) model has been the subject of 
intensive theoretical work
\cite{Nag74,Nol79a,Ovc91}.  The model describes the
exchange coupling of itinerant band electrons to localised magnetic
moments. The model is applicable to magnetic semiconductors
like the europium chalcogenides \cite{Wachter79} EuX (X=O, S, Se, Te)
and the chromium chalcogenide spinels \cite{Haas70} MCr$_2$Y$_4$ (M=Hg,
Cd; Y=S, Se). It is also used to describe
metallic local moment systems such as Gd,
Tb, and Dy \cite{Legvold80}.
Many characteristics of these materials
may be explained by a correlation between the localised
``magnetic'' states (4f, e.~g.~) and extended conduction band states
(5d-6s, e.g.). In the s-f model this correlation is represented by an
intraatomic 
exchange interaction.

At present magnetic phenomena at surfaces, interfaces, and in thin
films attract broad attention both theoretically and experimentally 
due to the question of phase transitions in dimensionally reduced systems. 
In particular, the magnetism of thin Gd films has been the subject of 
intensive experimental effort.  
In contrast to 3d-metals, Gd surfaces seem to have
a Curie temperature which is larger than the $T_C$ value of the bulk 
material \cite{RE86}. Similar
surface enhanced magnetism has been observed in Tb \cite{RJR88}. Wu et
al.~\cite{Wu91} have modelled an antiferromagnetic coupling between
the surface layer and the bulk. First experimental observations were made using
spin-polarised low-energy electron diffraction (LEED) and
photoemission spectroscopy (PES) \cite{WAG+85}. Further spinpolarised
photoemission 
studies \cite{MGE92,TWW+93,SNBK93,VRC93} did not confirm the original
observation and indicated a ferromagnetic coupling between surface and bulk.
Farle et al.~ \cite{FBS+93} measured the Curie-temperature of
layer-by-layer-grown Gd(0001) films as a function of film thickness. 
Donath et al.~\cite{DGP96} using spin resolved inverse
photoemission did not find any indication for exceptional
surface magnetic properties such as an enhanced Curie-temperature or
magnetic order at the surface which is different from bulk.

In particular, the behaviour of a Gd (0001) surface state has been discussed 
controversy.
Federov  et al.\cite{FSK94} and Wesch\-ke  et al.\cite{WSLM+96}  
find a Stoner-like temperature dependence of the
exchange splitting for the Gd (0001) surface. For the strained Gd(0001) Waldfried
et al.~ \cite{WWMVD97} observe a wave vector dependent exchange splitting.
They found the electronic structure at the surface to be different from that
of bulk and a substantially increased Curie temperature at the surface.
The FLAPW-calculations performed by Wu et al.\ \cite{WLF91} show a surface
state near the $\bar{\Gamma}$ point and an enhanced magnetic moment of
the Gd (0001) surface.

It is not only the dimensionally reduced
Gd which is of interest here, even bulk Gd is far from being completely
understood.  In an earlier study Nolting et al. \cite{NDB94} have observed that
the a priori non-magnetic (5d, 6s)-conduction and valence bands
exhibit a marked non uniform magnetic response which depends on the
positions within the Brillouin zone and on the subband. 
This may be the reason for the fact that the experimental situation is
controversial. 
Kim et al.~\cite{KAEKH94} found a $T$-dependent spin splitting of
occupied conduction electron states, which collapse in a Stoner-like fashion
for $T\rightarrow T_{\rm C}$. From photoemission experiments 
Li et al.~ \cite{LZDO92,LZDWO92} conclude
that the exchange splitting must be wave-vector
dependent, collapsing for some $\vec{k}$ values, while for others no
collapse occurs as a function of increasing
temperature. This fairly complicated behaviour of the exchange splitting in the
bulk material must be expected for Gd-films, too.

It is a challenge to perform an electronic structure calculation
for a local-moment ferromagnet of reduced dimension in such a manner as to
realistically incorporate correlation effects. 
In our previous paper \cite{SMN96} we 
proposed a simplified model which should be applicable to a local
moment film of finite thickness. The special situation is considered of
a single electron in an otherwise empty conduction band and coupled to the
ferromagnetically saturated local spins of a simple cubic (s.c.) film. This
model is applicable to a film of a ferromagnetic semiconductor such as EuO, EuS
at $T=0 {\rm K}$. This limiting case can be solved exactly for the bulk
\cite{AE82,NoDu85,NMJR96}. Its significance arises from the fact that 
all relevant correlation effects which are either found or expected to occur
for finite band occupations and arbitrary temperatures \cite{NDB94,NRMJ96}, 
do already appear in this simplified but tractable special model.
In our previous paper \cite{SMN96} an exact solution of the model 
has been given for a film of finite thickness
 showing the interplay between structure and many-body effects.
The film was formed by cutting a slab from the bulk material and
leaving all bulk
properties (such as intraatomic hopping) unmodified within the whole film.
However, this assumption is somewhat unrealistic.
Due to a wide range of physical effects near surfaces surface properties may
be significantly different from bulk properties
\cite{RE86,RJR88,WAG+85,TWW+93,CottTill,WD72}. 

In this work the variation of the hopping integrals is considered within 
the surface layer and between the surface layer 
and the layer nearest to the surface layer. We will demonstrate the modifying
influence of the surface manifesting itself in the appearance of Tamm surface
states \cite{Tam32}. 
We perform the model calculations to understand qualitatively  the surface
influence  an a local moment film.

In the next section we briefly recall the model and its exact solution
\cite{SMN96} and then show how the description can be extended to 
incorporate the variation of near surface hopping
integrals. For a film of finite thickness the results of  numerical
calculations are presented  in Section \ref{sec:results}, followed by a
summary and an outlook.

In this paper no concrete substance is focused on to as we perform 
the model calculations to qualitatively understand the surface
influence of the surface on the electronic properties of the local moment film.

\section{Theoretical Model}               
We investigate a film with a simple cubic structure. The film is obtained by
stacking $n$
layers parallel to the (100)-plane of the s.~c. crystal. Each lattice
point of the film is indicated by a  Greek
letter $\alpha, \beta, \gamma, \dots$ denoting the layer index
and an index $i, j, k, \dots$
numbering the sites within a given layer. Each layer possesses two-dimensional
symmetry, i.~e.~ the
thermodynamic average of any site-dependent operator
$\tens{O}_{i\alpha}$ depends only on the layer index $\alpha$.

We use the s-f (s-d) model as it is believed to yield a good
description for 
local-moment semiconductors and metals \cite{Nag74,Nol79a,Ovc91}.  The
complete model-Hamiltonian
  \begin{equation}
    \label{Hamilton}
    \tens{H}=\tens{H}_{\rm s}+\tens{H}_{\rm f}+\tens{H}_{\rm sf}
  \end{equation}
  consists of three parts.
The first
  \begin{equation}
    \tens{H}_{\rm s}=\sum_{ij\alpha\beta} \left(
      T_{ij}^{\alpha\beta}-\mu\delta_{ij}^{\alpha\beta} \right)
    \tens{c}^{+}_{i\alpha\sigma} \tens{c}_{j\beta\sigma}
  \end{equation}
  describes the itinerant conduction electrons as s-electrons.
  $\tens{c}^{+}_{i\alpha\sigma}$ and $\tens{c}_{i\alpha\sigma}$
  are, respectively, the creation and annihilation operators of an
  electron with the spin $\sigma (\sigma=\uparrow, \downarrow)$
  at site $\tens{\vec{R}}_{i\alpha}$.  $T_{ij}$ is the hopping
  integral.

  Each lattice site $\tens{\vec{R}}_{i\alpha}$ is occupied
  by a localised magnetic moment, represented by a spin operator
  $\tens{\vec{S}}_{i\alpha}$. The exchange coupling between these
  localised moments is expressed by the Heisenberg
  Hamiltonian:
  \begin{equation}
    \tens{H}_{\rm f}=\sum_{ij\alpha\beta} J_{ij}^{\alpha\beta}
    \tens{\vec{S}}_{i\alpha} \tens{\vec{S}}_{j\beta}
  \end{equation}
  $J_{ij}^{\alpha\beta}$ are exchange integrals.\\  
  The interaction of electrons and localised spins occurs via an
  intraatomic exchange. It gives rise to the third term in Eq.~(\ref{Hamilton}):
  \begin{equation}
    \tens{H}_{\rm sf}=-J\sum_{i\alpha} \tens{\vec{S}}_{i\alpha}
    \tens{\sigma}_{i\alpha}
  \end{equation}
  $J$ is the s-f exchange coupling constant and
  $\tens{\sigma}_{i\alpha}$ is the Pauli spin operator of the
  conduction band electrons. Using the second-quantised form of
  $\sigma_{i\alpha}$ and the abbreviations
  \begin{equation}
    \label{second_quant}
    \tens{S}^{\sigma}_{j\alpha}=\tens{S}^x_{j\alpha}+ {\rm
      i}z_{\sigma}\tens{S}^y_{j\alpha} ;\:\:z_{\uparrow}=+1,
    \:\:z_{\downarrow}=-1, 
  \end{equation}
  $\tens{H}_{\rm sf}$ can be written as
  \begin{equation}
    \label{H_sf2q}
    \tens{H}_{\rm sf}=-\frac{1}{2}J\sum_{i\alpha\sigma}\left(
    z_{\sigma}\tens{S}^z_{i\alpha}\tens{n}_{i\alpha\sigma}+
    \tens{S}^{\sigma}_{i\alpha}\tens{c}^+_{i\alpha -\sigma}
    \tens{c}_{i\alpha\sigma}\right).
 \end{equation}

 All physically relevant information of the system can be derived from the
 retarded single electron Green 
 function: 
 \begin{eqnarray}
   \label{seGF}
   G^{\alpha\beta}_{ij\sigma}(E)&=&
   \left\langle\left\langle c_{i\alpha\sigma}\:;\:
   c^+_{j\beta\sigma}\right\rangle\right\rangle_{E}\nonumber\\
   &=&-{\rm i}\int\limits_0^{\infty}{\rm d}t\:{\rm e}^{-\frac{\rm
     i}{\hslash}Et}
 \left\langle\left[ 
 \tens{c}_{i\alpha\sigma}(t),\tens{c}^+_{i\beta\sigma}(0)
  \right]_+\right\rangle.          
\end{eqnarray}
Here $[\dots,\dots]_+$ is the anticommutator. The creation and
annihilation operators are used in their (time dependent) Heisenberg
representation. We perform a Fourier transformation with\-in the layer
\begin{equation}
  \label{layGF}
  G_{\vec{k}\sigma}^{\alpha\beta}(E)=\frac{1}{N}
  \sum_{ij}{\rm e}^{{\rm
      i}\vec{k}(\vec{R}_i-\vec{R}_j)}
  G_{ij\sigma}^{\alpha\beta} 
\end{equation}
Eq.~(\ref{layGF}) conforms to the two-dimensional translation symmetry. 
$N$ is the number of 
sites per layer, $\vec{k}$ a wave-vector from the first two-dimensional
Brillouin zone. 
Another relevant quantity is the spectral density, defined as
the imaginary part of the Green function
\begin{equation}
  \label{Spektral}
  S^{\alpha\beta}_{\vec{k}\sigma}(E)=-\frac{1}{\pi}\Im
  G_{\vec{k}\sigma}^{\alpha\beta}(E+{\rm i}0^+)\;.
\end{equation}
$S^{\alpha\beta}_{\vec{k}\sigma}$ is directly connected to observable
quantities within angle and
spin resolved (inverse) photoemission experiments. The 
wave-vector summation of the spectral density Eq.~(\ref{Spektral}) yields a
layer-dependent (local) quasiparticle density of states,
\begin{equation}
  \label{LDOS}
  \rho_{\alpha\sigma}(E)=\frac{1}{\hslash N}\sum_{\vec{k}}
  S^{\alpha\alpha}_{\vec{k}\sigma}(E)=
  \frac{1}{\hslash}S^{\alpha\alpha}_{ii\sigma}(E) 
\end{equation}
which will play an important role in following discussion.

The many-body problem posed by the Hamiltonian Eq.~(\ref{Hamilton}) is non
trivial. Up to now no analytic solution is found even for the bulk. The model
is 
widely used and has proved to be realistic for applications to
semiconducting or metallic local moment ferro- and antiferromagnets
\cite{Nol79a,Ovc91,Wachter79,NDB94}. We investigate the special
situation of a thin  saturated ferromagnetically ordered film, i.\ e.\
$\langle\tens{S}^z_{i\alpha}\rangle=S$, with a single electron in an otherwise
empty conduction band at zero-temperature ($n=0, T=0$) and interacting via the
s-f exchange 
Eq.~(\ref{H_sf2q})  with the localised spin system of the film.

The solution of the many-body problem is described very briefly. Details can
be found in our previous paper \cite{SMN96}. The 
equation of motion of the single electron Green function Eq.~(\ref{seGF}) can be
written 
\begin{equation}
  \label{BewegI}
  EG^{\alpha\beta}_{ij\sigma}=\hslash\delta^{\alpha\beta}_{ij} +
  \sum_{l\gamma}T^{\alpha\gamma}_{il} G^{\gamma\beta}_{lj\sigma} -
  \frac{J}{2} \left( z_{\sigma}\Gamma^{\alpha\alpha\beta}_{iij\sigma} +
  F^{\alpha\alpha\beta}_{iij\sigma} \right)
\end{equation}
where we have introduced two ``higher'' Green functions
\begin{eqnarray}
  \label{Ising}
  \Gamma^{\alpha\beta\gamma}_{ijk}&=&
  \left\langle\left\langle \tens{S}^z_{i\alpha}\tens{c}_{j\beta\sigma}\:;\:
  \tens{c}^+_{k\gamma\sigma}\right\rangle\right\rangle_{E}
  \stackrel{\scriptstyle n\rightarrow0, T\rightarrow0
    }{-\!\!\!-\!\!\!-\!\!\!-\!\!\!\longrightarrow }\hslash S
  G^{\beta\gamma}_{jk\sigma} ,\\ 
  \label{Flip}
   F^{\alpha\beta\gamma}_{ijk}&=&
   \left\langle\left\langle \tens{S}^{-\sigma}_{i\alpha}\tens{c}_{j\beta-\sigma}
   \:;\:
   \tens{c}^+_{k\gamma\sigma}\right\rangle\right\rangle_{E}\;.
\end{eqnarray}
For the case of a completely saturated local moment system the function
Eq.~(\ref{Ising}) can be expressed as a product of $\hslash S$ and the
single-electron Green function Eq.~(\ref{seGF}). This arises because in the Green
functions Eqs.~(\ref{Ising}) and (\ref{Flip}) the averaging has to be done using the
magnon and electron vacuum
$\left(\sum_{\sigma}\langle\tens{n}_{\alpha\sigma}\rangle\equiv0\right)$. 
Eq.~(\ref{Flip}) can be interpreted as a function expressing ``spin flips'' at
certain lattice sites. Applying a two-dimensional Fourier transformation to
the equation of motion Eq.~(\ref{BewegI}) yields 
\begin{eqnarray}
  \label{BewegII}
  \lefteqn{\left(E+\textstyle\frac{1}{2}J\hslash S \right)
  G^{\alpha\beta}_{\vec{k}\sigma}}\hspace{3em}\nonumber\\
  &=& \hslash\delta_{\alpha\beta} + \sum_{\gamma}
  T^{\alpha\gamma}_{\vec{k}}
  G^{\gamma\beta}_{\vec{k}\sigma} - \frac{J}{2\sqrt{N}}
  \sum_{\vec{q}}
  F^{\alpha\alpha\beta}_{\vec{k}\vec{q}\sigma} 
\end{eqnarray}
where $\vec{k}$ and $\vec{q}$ are wave vectors of the two
dimensional Brillouin zone and 
\begin{equation}
  \label{hopp_k}
  T^{\alpha\beta}_{\vec{k}}= \frac{1}{N}\sum_{ij} 
  {\rm e}^{{\rm i} \vec{k}(\vec{R}_i-\vec{R}_j)} \:
  T^{\alpha\beta}_{ij} 
\end{equation}
the energy is in Bloch presentation.

In order to solve Eq.~(\ref{BewegII}) we have to evaluate the ``spin flip''
function 
$F^{\alpha\gamma\beta}_{\vec{k}\vec{q}\sigma}$. If the spin
$\sigma$ of the single electron points into the same direction as the
ferromagnetically saturated lattice ($\sigma=\uparrow$) the electron cannot
exchange its spin with the local moment system. This corresponds to the
disappearance of the ``spinflip'' function 
Eq.~(\ref{Flip}),
\begin{equation}
  \label{Flip=0}
  F^{\alpha\alpha\beta}_{iij\uparrow}
  \stackrel{\scriptstyle n\rightarrow0, T\rightarrow0
    }{-\!\!\!-\!\!\!-\!\!\!-\!\!\!\longrightarrow }0\;.
\end{equation}

The case of a $\sigma=\downarrow$ electron is more complex. There are
many possibilities to exchange its spin with the local moment
system.
As a consequence the ``spin flip'' function Eq.~(\ref{Flip}) does not 
vanish as in the $\uparrow$-case. Nevertheless it turns out that the equation
of motion for 
$F^{\alpha\alpha\beta}_{iij\downarrow}$ decouples exactly.
As a result the ``spin-flip'' function can be expressed in terms of the single
electron Green function (for details see \cite{SMN96})
\begin{equation}
  \label{exp_F}
  -\frac{J}{2\sqrt{N}}\sum_{\vec{q}}
  F^{\alpha\alpha\beta}_{\vec{k}\vec{q}\sigma} =
  C_{\alpha} G^{\alpha\beta}_{\vec{k}\downarrow}
\end{equation}
where
\begin{equation}
  \label{def_C}
  C_{\alpha}= \frac{\frac{1}{2}J^2\hslash^2SB_{\alpha}}{1-\frac{1}{2}J\hslash
  B_{\alpha}},
\end{equation}
\begin{equation}
  \label{def_B}
    B_{\alpha}=\frac{1}{N}\sum_{\vec{q}}\left(
    A^{-1}_{\vec{q}\downarrow} \right)^{\alpha\alpha},
\end{equation}
and
\begin{equation}
  \label{def_matrix}
  \left(A_{\vec{k}\sigma}(E)\right)^{\alpha\beta}= -
  T^{\alpha\beta}_{\vec{k}} + \delta^{\alpha\beta} \left( E +
      \textstyle \frac{1}{2}z_{\sigma}J\hslash S\right)\;.
\end{equation} 

Combining Eqs.~(\ref{BewegII}), (\ref{Flip=0}), and (\ref{exp_F}) one obtains
\begin{equation}
  \label{Formal_Loesung}
  \tens{G}_{\vec{k}\sigma}=\hslash\tens{D}^{-1}_{\vec{k}\sigma}
\end{equation}
where 
\begin{equation}
  \label{D_Matrix}
  \left(\tens{D}_{\vec{k}\sigma}\right)^{\alpha\beta}=
  \left(\tens{A}_{\vec{k}\sigma}\right)^{\alpha\beta}-
  \delta^{\alpha\beta}_{\downarrow\sigma}C_{\alpha}\;.
\end{equation}

\begin{figure}[htbp]
  \begin{center}
    \mbox{}
    \epsfig{file=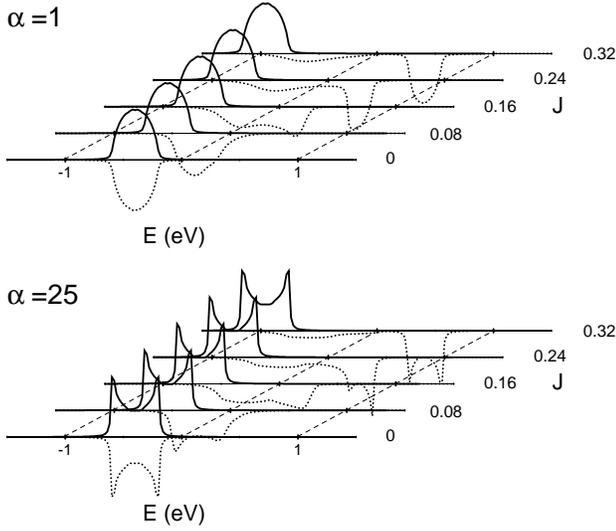,width=0.95\linewidth}\\\mbox{}
    \caption{Local spin-$\sigma$ spectral densities of the first
    and the 25th layer of a 50-layer s.c.-(100)-film with uniform hopping
    ($\epsilon_{\parallel} = \epsilon_{\perp} = 1$)
    for different s-f exchange constants $J=0, 0.08, 0.16, 0.24, 0.32$.
    The spectra for the spin-$\uparrow$ and the spin-$\downarrow$ electron
    are drawn as a solid line in positive direction and as a dotted line in
    negative direction, respectively}
\label{fig:topcenter}\end{center}
\end{figure}

To start our discussion we recall the basic features of the 
spectra of a single spin-$\sigma$ electron as presented in our previous paper
\cite{SMN96}. Fig.~\ref{fig:topcenter} shows the spectral densities of both 
spin-$\uparrow$ and spin-$\downarrow$ electron at
$\vec{k}=\bar{\Gamma}$ 
for the first and the centre layer
of a 50-layer film with uniform hopping.
For vanishing s-f interaction $J$ there is no way to distinguish between
a spin-$\uparrow$ and a spin-$\downarrow$ electron. The spectra therefore
coincide.  
If the s-f interaction $J$ is switched on the spectrum of the spin-$\uparrow$ 
electron is rigidly shifted by a
constant energy of $-\frac{1}{2} JS$, since the electron has no chance to
exchange its spin with the perfectly aligned local-moment system. 
However, this is possible for the spin-$\downarrow$ electron. 
For small exchange coupling $J$ a 
slight deformation of the
free spin-$\downarrow$ spectral density sets in. For intermediate and strong
s-f-interaction the spectrum splits into two parts. The higher energetic part
represents a polarisation of the immediate spin
neighbourhood of the electron due to a repeated emission and reabsorption
of magnons. The result is a polaron-like quasiparticle called the ``magnetic
polaron''. The low-energetic part of the spectrum is a scattering band which
corresponds to the emission
of a magnon by the spin-$\downarrow$ electron without reabsorption, but with a
spin-flip of the electron. For details of the discussion see also \cite{SMN96}.

\begin{figure}[bh]
  \begin{center}
    \mbox{}
    \epsfig{file=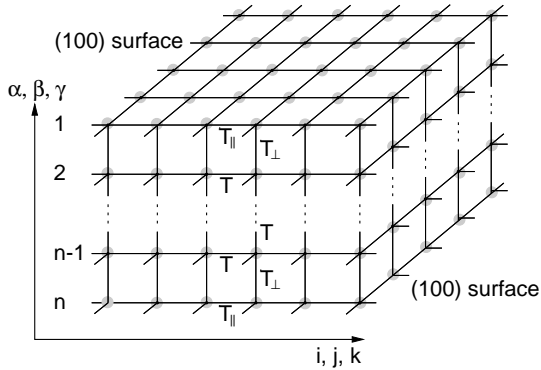,width=.8\linewidth}
    \caption{Model structure of an 100-layer film with s.~c.~
      structure. Nearest neighbour hopping $T_{\parallel}$: within the surface
      layer; $T_{\perp}$: between the surface and the layer nearest to the
      surface layer; $T$: 
      within and between inner layers}
    \label{fig:layer}
  \end{center}
\end{figure}

In contrast to \cite{SMN96} we are here interested in surface states.
We apply the above theory to a s.~c.~film consisting of $n$ layers oriented
parallel to the (100)-surface as drawn schematically in Fig.~\ref{fig:layer}.
We restrict the electron hopping to nearest neighbours only i.\ e.\ 
\begin{equation}
  \label{hopp_nn}
  T^{\alpha\beta}_{ij}=\delta^{\alpha\beta}_{i,j\pm\Delta}T^{\alpha\alpha}
  +\delta^{\alpha,\beta\pm1}_{ij}T^{\alpha\beta} 
\end{equation}
were
$\Delta=(0,1), (0,\bar{1}), (1,0), (\bar{1},0)$. 
$T^{\alpha\beta}$ is the hopping between the layers $\alpha$ and
$\beta=\alpha\pm1$ 
and $T^{\alpha\alpha}$ within the layer $\alpha$. 
In order to study surface states we vary
the hopping within the surface $T_{\parallel}=\epsilon_{\parallel} T$ and
between the 
surface and the layer nearest to the surface layer $T_{\perp}=\epsilon_{\perp}
T$ by introducing 
the parameters $\epsilon_{\parallel}$ and $\epsilon_{\perp}$
\begin{equation}
  \label{hopp_matrix}
  T^{\alpha\beta}=\left(
    \begin{array}{cccccc}
      \epsilon_{\parallel} T &
      \epsilon_{\perp}     T &
      0 & &
      \cdots &
      0  \\
      \epsilon_{\perp}     T &
      T &
      T & & &
      \vdots  \\
      0 &
      T & &
      \ddots &  &   \\
      &  &
      \ddots &  &
      T &
      0  \\
      \vdots &  & &
      T &
      T &
      \epsilon_{\perp} T   \\
      0 &
      \cdots &  &
      0 &
      \epsilon_{\perp} T &
      \epsilon_{\parallel} T
    \end{array}
    \right)
\end{equation}
in contrast to a uniform hopping $T$ in our previous paper
\cite{SMN96}.
Here $\epsilon_{\parallel}$ and $\epsilon_{\perp}$ are considered as 
model parameters.
In the numerical calculations the bulk hopping is set to $T=-0.1 {\rm
  eV}$. The investigation of the influence of modified 
surface hopping is performed for the example of an infinite (100)-layer
s.~c.~film. The
elements of the Bloch matrix are defined in Eq.~(\ref{hopp_k}) where the summation
includes nearest neighbours only. Thus the diagonal elements of the Bloch
matrix are given by 
\begin{equation}
  \renewcommand{\arraystretch}{1.7}
  \begin{array}{lccl}
  \label{hopp_diag}
    T^{11}_{\vec{k}}&=&
    2\epsilon_{\parallel}&T(\cos(k_x a) + \cos(k_y a)), \\
    T^{\alpha\alpha}_{\vec{k}}&=&
    2&T(\cos(k_x a) + \cos(k_y a)),
  \end{array}
\end{equation}
and the first upper and lower diagonal elements by
\begin{eqnarray}
  \label{hopp_off}
  \renewcommand{\arraystretch}{1.7}
  \begin{array}{lclccl}
    T^{12}_{\vec{k}}&=&
    T^{21}_{\vec{k}}&=&
    \epsilon_{\perp}T,\\
    T^{\alpha,\alpha+1}_{\vec{k}}&=&
    T^{\alpha+1,\alpha}_{\vec{k}}&=& 
    T.
  \end{array}
\end{eqnarray}

For either a semi-infinite system or a sufficiently thick film some essential 
analytical predictions can be made which are subject of a forthcoming paper
\cite{MSN97}. 
For the $\uparrow$-electron the layer dependent retarded single electron 
Green function
$G^{\alpha\beta}_{\vec{k}\uparrow}(E)$ can be calculated analytically
and both the existence and the behaviour of surface states can be studied.

Modifying the hopping within the first  layer by more than $25\%$ i.~e.
by a factor $\epsilon_{\parallel}\le\frac{3}{4}$  or 
$\epsilon_{\parallel}\ge\frac{5}{4}$ and keeping all the other hopping
integrals unchanged results in a single surface state which splits off, 
at the lower or upper edge of the bulk band. 
This surface state first emerges  for 
$\vec{k}=\bar{\Gamma}$, \mbox{$\overline{\rm M}$} 
from the bulk band and from there it spreads
for larger modification to the rest of Brillouin zone.

If the hopping within the whole film remains constant but 
the hopping between the first 
and the second layer is changed by a factor $\epsilon_{\perp}\ge\sqrt{2}$ then
two surface states split off one on each side of the bulk band.
The splitting is $\vec{k}$ independent.

It proves that the maximum of the spectral weight of surface states is located 
either at
the first or at the second layer. From there it  drops exponentially as a
function of the distance from the surface. The closer the surface state is
located to bulk states the slower the descent.

Generally the correlation is observed such that the more strong\-ly modified
the hopping ($|\epsilon_{\parallel}-1|,\epsilon_{\perp}-1$) becomes the larger 
gets
the spectral weight  of surface states of the layers close to the surface.

\begin{figure}[htbp]
  \begin{center}
    \mbox{}
    \epsfig{file=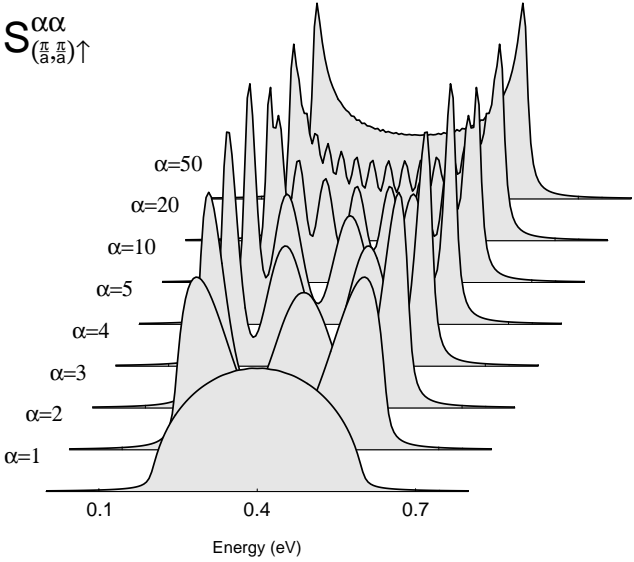,width=0.95\linewidth}
    \caption{Local spin-$\sigma$ spectral densities ($J=0,
      \sigma=\uparrow,\downarrow$) of the layer ($\alpha=1, 2, 3, 4, 5, 10, 20,
      50$) of a 100-layer film at
      $\vec{k}=(\frac{\pi}{a},\frac{\pi}{a})$ as a function of
      energy (eV). The hopping is uniform within the film:
      $\epsilon_{\parallel}=\epsilon_{\perp}=1$ 
      }
    \label{fig:uniform}
  \end{center}
\end{figure}

\begin{figure}[htbp]
  \begin{center}
  \mbox{}
    \epsfig{file=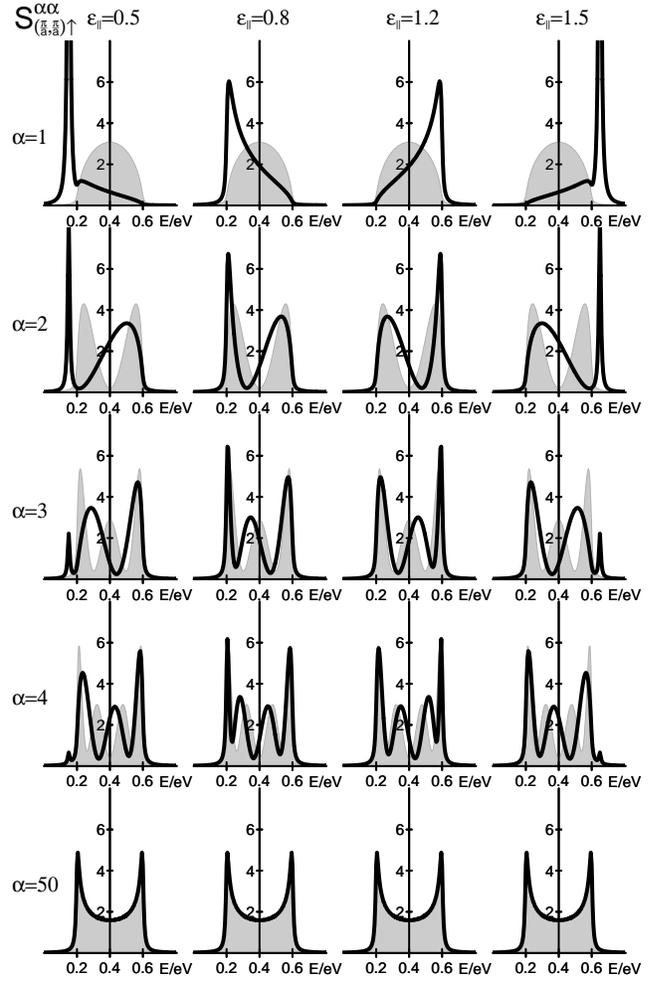,width=0.95\linewidth}\\\mbox{}
    \caption{Local spin-$\uparrow$ spectral densities ($J=0,
      \sigma=\uparrow,\downarrow$) of the layer ($\alpha=1, 2, 3, 4, 50$) of a
      100-layer film at $\vec{k}=(\frac{\pi}{a},\frac{\pi}{a})$ as a
      function of energy 
    (eV) for modified hopping within the first layer by the factor
    $\epsilon_{\parallel}=0.5, 0.8, 1.2, 1.5$ and constant hopping between the
    first and the second layer
    ($\epsilon_{\perp} = 1$) (black line). The grey background shows the
    spectral density for the case of uniform hopping}
\label{fig:eps_para}\end{center}
\end{figure}
\section{Results}
\label{sec:results}%
We discuss our results in terms of the spectral density
$S^{\alpha\alpha}_{\vec{k}\sigma}(E)$ as defined in Eq.~(\ref{Spektral}), and
the quasiparticle density of states $\rho_{\alpha\sigma}(E)$ Eq.~(\ref{LDOS}).

\subsection{Spin-$\tens{\uparrow}$-electron}
\label{subsec:up}
For the $\sigma=\uparrow$-electron the s-f-exchange $J$ 
results only in a rigid shift
of the spectrum (\ref{def_matrix}-\ref{D_Matrix}). Therefore we can choose $J=0$ to study the influence
of modified hopping.             
Fig. \ref{fig:uniform} shows the $\sigma$-spectral density 
$S^{\alpha\alpha}_{\vec{k}\sigma}$ for the $\alpha$-th layer of a 
100-layer film at 
$\vec{k}=(\frac{\pi}{a},\frac{\pi}{a})$ for uniform hopping
($\epsilon_{\parallel}=\epsilon_{\perp}=1$) and the s-f interaction
$J=0$ switched-off.
\begin{figure}[htbp]
  \begin{center}
  \mbox{}
    \epsfig{file=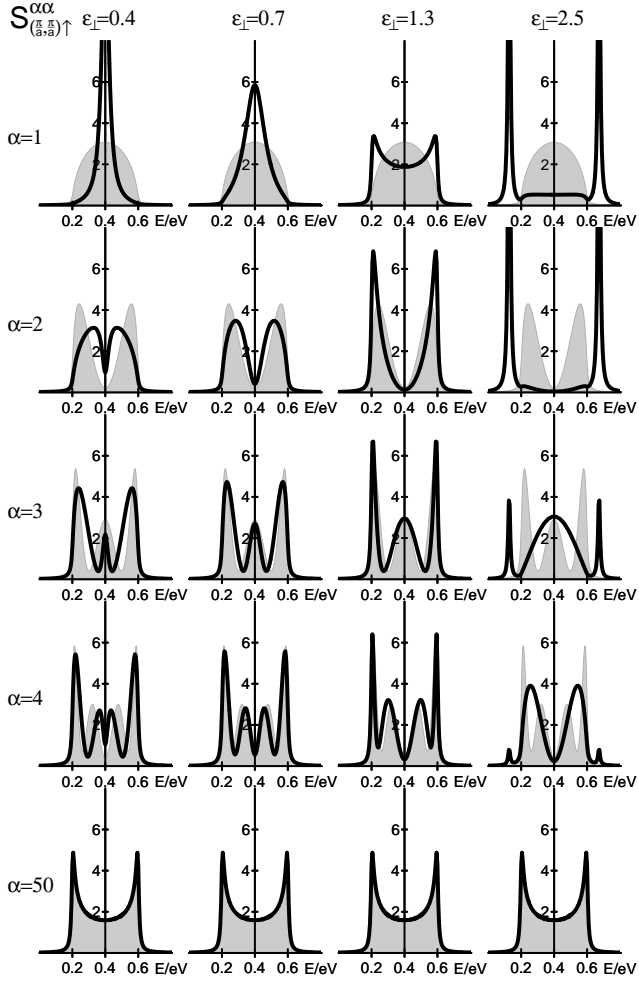,width=0.95\linewidth}\\\mbox{}
    \caption{Local spin-$\uparrow$ spectral densities ($J=0,
      \sigma=\uparrow,\downarrow$) of the layer ($\alpha=1, 2, 3, 4, 50$) of a
      100-layer film at $\vec{k}=(\frac{\pi}{a},\frac{\pi}{a})$ as a function of energy
    (eV) for modified hopping between the first and second layer by the factors
    $\epsilon_{\perp}=0.4, 0.7, 1.3, 2.5$ and constant hopping within the
    first layer
    ($\epsilon_{\parallel} = 1$) (black line). The grey background is
    the spectral density for the case of uniform hopping}
\label{fig:eps_perp}\end{center}
\end{figure}
The spectral densities vary from layer to layer as a consequence of the broken
symmetry at the surface. For each $\vec{k}$ point the local spectral density
of a given layer equals
the density of states of an atom in a one-dimensional finite tight-binding linear
chain. 
For the given situation at the \mbox{$\overline{\rm M}$} point of the two dimensional Brillouin zone
the centre of gravity for each layer is given by
$T^{\alpha\alpha}(\frac{\pi}{a},\frac{\pi}{a})=0.4{\rm  eV}$ and the total bandwidth is
$-4T^{\alpha,\alpha+1}=0.4{\rm  eV}$. The lower and upper band edges are
$0.2{\rm  eV}$ and $0.6{\rm  eV}$ respectively. For the innermost layers
($\alpha=50, 51$) the spectral density approaches the one dimensional
tight-binding 
density of states whereas for the surface layers ($\alpha=1, 100$) it
approaches the semi-elliptic one. 

The variation of hopping is done by either varying 
$\epsilon_{\parallel}$ or $\epsilon_{\perp}$ while keeping the
other fixed to determine the influence of both parameters.

Fig.\ \ref{fig:eps_para} exhibits the layer dependent local spin-$\uparrow$
spectral density for the same wave vector $\vec{k}=(\frac{\pi}{a},\frac{\pi}{a})$ as in
Fig.\ \ref{fig:uniform}. The hopping is modified within the surface layer
($\epsilon_{\parallel}\neq1$) while 
$\epsilon_{\perp}=1$. Deviations of intra layer hopping in the surface
layer result in a transfer of spectral weight for $\epsilon_{\parallel}<1$
towards lower and for $\epsilon_{\parallel}>1$ towards higher energies,
respectively. The transfer is most significant for the surface layers and
decreases towards the bulk. For the inner (bulk like) layers the modification
of 
hopping at the surface layers has no effect. If the difference between bulk
and surface hopping is strong enough the transfer of spectral weight causes
the splitting off of a surface $\delta$-like peak. This happens for
$\epsilon_{\parallel}<0.75$ at the lower and for $\epsilon_{\parallel}>1.25$ at the
upper band edge, respectively. The features observed for reduction of hopping
are symmetric to those observed for an increase of the hopping matrix element.

We obtain the curves shown in Fig.\ \ref{fig:eps_perp} by varying the hopping
between the surface layer and the layer nearest to the surface layer 
and leaving
all other hopping integrals within a 100-layer film
unchanged. If the hopping between the
first and the second layer decreases ($0\le\epsilon_{\perp}<1$) 
spectral weight is symmetrically transferred from band edges towards 
the centre of the band for the spectral density. In the limiting case
$\epsilon_{\perp}\rightarrow 0$ the spectral density of the first layer is
converted to a $\delta$-peak whereas the spectral density
of the second layer $\alpha=2$ approaches the shape of the surface spectral
density of a film with uniform hopping. The reason is that
$\epsilon_{\perp}=0$ means a complete decoupling of the first layer from the
rest of the film inasmuch the second layer plays the role of a surface.

\begin{figure}[thb]
  \begin{center}
  \mbox{}
    \epsfig{file=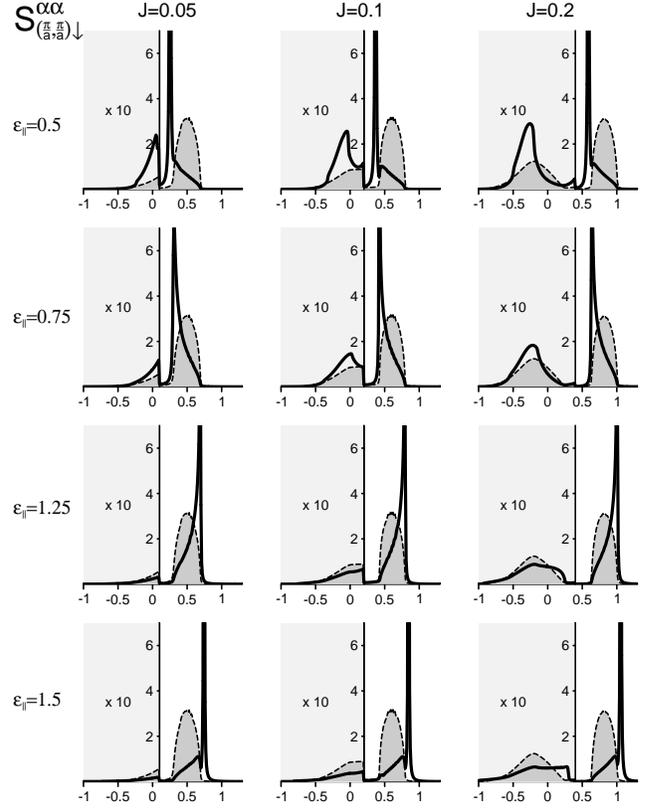,width=0.95\linewidth}\\\mbox{}
    \caption{Local spin-$\downarrow$ spectral density of the surface layer 
    ($\alpha=1$) of a 100 layer film at 
    $\vec{k}=(\frac{\pi}{a},\frac{\pi}{a})$ with modified hopping
    within the first layer ($\epsilon_{\parallel}=0.5, 0.75, 1.25$) while
    $\epsilon_{\perp} = 1$ for different
    s-f exchange coupling constants $J=0.05, 0.1, 0.2$. The dashed lines show
    the spectral density for the case of uniform hopping. 
    The values on the low energy side of the spectra (light
    grey background) are multiplied by the factor 10}
\label{fig:down_para}\end{center}
\end{figure}
If the hopping is augmented, as expressed by $\epsilon_{\perp}>1$, the spectral
weight is symmetrically distributed towards the edges of the bulk band. 
For a sufficiently large modification  of the hopping we
observe the simultaneous splitting off of two surface states, one at the lower 
and the other at the upper edge of the bulk band. These surface states can be
observed for the spectral densities ($\epsilon_{\perp}=2.5$) of the first four
layers ($\alpha=1, 2, 3, 4$). The spectral weight of these states decreases
exponentially with the distance from the surface layer. The energy position of
the excitation is independent of the layer index $\alpha$. An analytical
investigation of the limits 
in parameter space $(\epsilon_{\perp},\epsilon_{\parallel})$, separating
regions with one two or without surface states, will be given in \cite{MSN97}.

\subsection{Spin-$\tens{\downarrow}$-electron}

\begin{figure}[thb]
  \begin{center}
  \mbox{}
    \epsfig{file=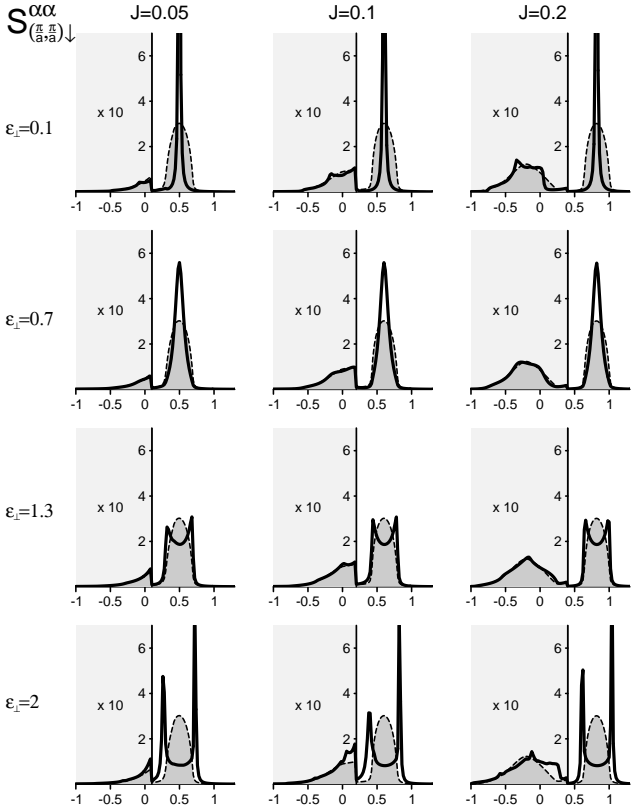,width=0.95\linewidth}\\\mbox{}
    \caption{Local spin-$\downarrow$ spectral density of the surface layer
    ($\alpha=1$) of
    a 100-layer film at $\vec{k}=(\frac{\pi}{a},\frac{\pi}{a})$ with modified
    hopping between the surface layer and the surface nearest layer
    ($\epsilon_{\perp}=0.25, 0.5, 2, 4$) and $\epsilon_{\parallel} = 1$ 
    for different
    s-f exchange coupling constants $J=0.05, 0.1, 0.2$. The dashed lines show
    the spectral density for uniform hopping. The values on the low energy 
    side of the spectra  (light grey background) are 
    multiplied by the factor 10}
\label{fig:down_perp}\end{center}
\end{figure}
The local spin-$\downarrow$ spectral density
of the surface layer of a 100-layer film at the \mbox{$\overline{\rm M}$} point 
($\frac{\pi}{a},\frac{\pi}{a}$) of the
Brillouin zone in case of different s-f exchange coupling constants $J$ can be
seen in Fig.~\ref{fig:down_para}. The hopping
is modified within the topmost layer. In analogy to the case of the
spin-$\uparrow$ electron  there is a splitting off of a surface state at the
 polaron band of the spectrum. 
This effect is independent on the s-f interaction $J$. The
scattering part located at lower energies is altered as a consequence of
modification of the spin-$\uparrow$ spectrum by $\epsilon_{\parallel}$. This
reflects the transfer of spectral weight  in the spin-$\uparrow$ spectral
density. However, a surface state does not show up at the scattering part of
the spectral density.

The same holds for the spin-$\downarrow$ spectral density of states for the 
situation of modified hopping between first and the second layer of a
100-layer film as can be seen in
Fig.~\ref{fig:down_perp}. Here also, the existence of surface states 
does only depend on the parameter $\epsilon_{\perp}$ 
and is independent on the s-f exchange
interaction $J$.
\begin{figure}[htbp]
  \begin{center}
  \mbox{}
    \epsfig{file=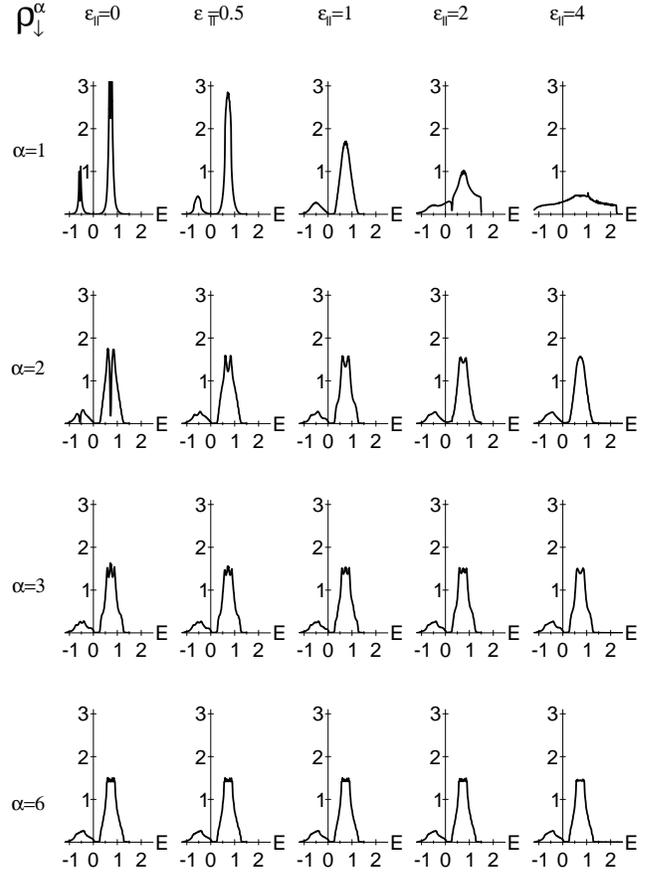,width=0.95\linewidth}\\\mbox{}
    \caption{Local spin-$\downarrow$ densities of states for
      $J=0.32$ of the layers 
      ($\alpha=1, 2, 3, 6$) of a 12 layer film with modified hopping within
      the first layer and $\epsilon_{\perp} = 1$}
\label{fig:dos_para}\end{center}
\end{figure}

Fig.~\ref{fig:dos_para} shows the local density of states (LDOS) of a 
12-layer film. The hopping within the surface layer is modified while all the 
other hopping integrals are left unchanged. 
For the surface layer ($\alpha=1$) we observe a drastic dependence of the LDOS
on the hopping $\epsilon_{\parallel}$. 
A reduction of the hopping within the first layer results in a narrowing of
the scattering
and the polaron part of the LDOS of the first layer. 
An increase of the hopping causes a broadening  of both  scattering
and polaron parts. For a substantial modification of the hopping
$\epsilon_{\parallel}>2$ both parts merge and the LDOS of the
second layer approaches the LDOS of the surface layer in the case
of uniform hopping.
One possible explanation is that if the hopping within the surface layer
becomes infinite ($\epsilon_{\parallel}\rightarrow\infty$)
 and the hopping between the surface layer and the surface
nearest layer remains finite the electron can only move within the surface
layer and will not jump to the second layer.
The consequence is an ``effective'' decoupling of the surface layer from the
rest of the film as in the case of ($\epsilon_{\perp}=0,
\epsilon_{\parallel}=1$) discussed in subsection \ref{subsec:up}.
The analytical solution \cite{MSN97} for $\sigma=\uparrow$ shows that in the
limit $\epsilon_{\parallel}\rightarrow\infty$ and $\epsilon_{\perp}=$const.~
the Green function of the second layer becomes equal to the Green function of
the surface layer in case of uniform hopping.
\begin{figure}[htbp]
  \begin{center}
  \mbox{}
    \epsfig{file=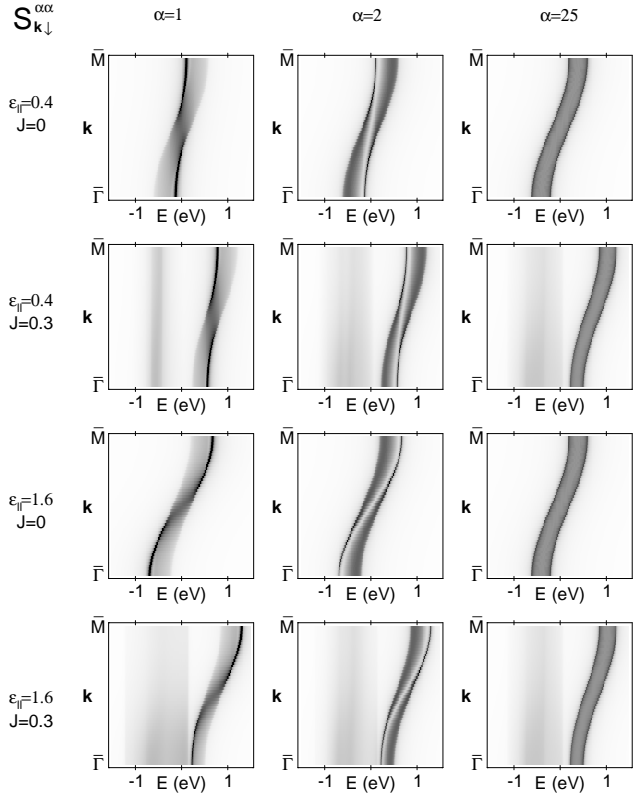,width=0.95\linewidth}\mbox{}
    \caption{Density plots of the local spin-$\downarrow$ spectral density of
      the 
      $\alpha=1, 2, 25$ layer of a 50 layer film as a function of energy and
      $\vec{k}\in(\overline{\Gamma},\mbox{$\overline{\rm M}$})$ for two s-f
      exchange coupling constants 
      $J=0, 0.3$ and modified hopping within the first layer
      ($\epsilon_{\parallel}=0.4, 1.6$). The hopping between all layers
      remains constant
      }
\label{fig:density}\end{center}
\end{figure}

Fig.~\ref{fig:density} offers a series of density plots of the local spectral
density of the first, second, and central layer of a 50 layer film as
a function of energy and wave vector $\vec{k}$ for different values of
the s-f exchange ($J=0, 0.3$) and modified hopping within the
first layer.
In the spectra of the first and second layer 
we observe the splitting off
of a surface state from the bulk band ($\alpha=25$) near the $\overline{\Gamma}$ and
near the 
\mbox{$\overline{\rm M}$} point of Brillouin zone. For $\epsilon_{\parallel}=0.4<1$
the splitting off 
takes place on the inner side of dispersion curve causing a narrowing of
polaron 
band of the LDOS 
for $\epsilon_{\parallel}=1.6>1$ on the outer side resulting in a 
broadening of it. 
As a consequence of the broadened or narrowed $\sigma=\uparrow$ LDOS (see
Fig.~\ref{fig:density}) for  $\epsilon_{\parallel}=1.6$ the scattering part is
broadened but for $\epsilon_{\parallel}=0.4$ it is narrowed.

\section{Summary} 

We have investigated the electronic quasiparticle spectrum of a ferromagnetic
local-moment film of finite thickness. A single band electron is coupled by an
intraatomic s-f interaction to the magnetic local-moment system. For 
a ferromagnetically saturated localised spin system, $T=0 \rm{K}$, an
exact solution of the problem was given in our previous work \cite{SMN96}.

In contrast to this, in this work
the investigation has been focused on the influence of modified hopping
integrals near the surface of the film. 
In tight binding approximation for the hopping integrals, 
either the hopping within the topmost layer or the hopping between the
first and the second layer has been modified. 

Generally the modification of the hopping causes a transfer of spectral weight
within the spectra.

For sufficiently modified hopping within the surface layer
($|\epsilon_{\parallel}-1|>\frac{1}{4}$), 
both for increased and decreased hopping and for both spin directions $\sigma$
of the single electron a surface state splits off from the
bulk band ($\sigma=\uparrow$) or the polaron band ($\sigma=\downarrow$),
respectively. The existence of surface states for a given
$\epsilon_{\parallel}$ depends on 
the wave vector $\vec{k}$ of the two dimensional Brillouin zone and starts at
the $\overline{\Gamma}$ and $\overline{\mbox{M}}$ point. 

{\sloppy If the hopping between the surface layer and
the sur\-face nearest layer $\epsilon_{\perp}$ is  ad\-e\-quately in\-creased two
sur\-face states show up, one at the lower edge and one at the up\-per edge of the
bulk spec\-trum.}

The spectral weight of the surface states exponentially
decays from the surface into the bulk.
The surface effects induced by modified hopping visible in the spin-$\uparrow$
spectrum are rendered by correlation in the spin-$\downarrow$ spectrum.

\begin{acknowledgement}
This work was supported by the Deutsche Forschungsgemeinschaft within the
Sonderforschungsbereich 290 (``Metallische d\"unne Filme: Struktur,
Magnetismus, und elektronische Eigenschaften''). One of us (R.~S.\ ) gratefully
acknowledges the support by the Studienstiftung des deutschen Volkes.
\end{acknowledgement}

\end{document}